\documentclass[12pt]{article}
\usepackage{graphicx}
\usepackage{amsfonts,amssymb,eucal,amsmath}
\begin{document}

 \parskip 0pt

\begin{center}{\bf \Large
%
Variation Principle for Calculation of Many-Particle Effects in
Crystals }
\\[0.2cm]
{\it  
Halina V. Grushevskaya$^{1}$ and Leonid I. Gurskii$^{2}$
}\\[0.2cm]
 {\it $^{1}$ Physics Department, Belarusian State University,\\
 4 Fr. Skorina Av., 220050 Minsk, BELARUS \\
 $^{2}$ Belarusian State
 University of Informatics  and Radio electronics,\\
 6 P. Brovka Str., 220027 Minsk, BELARUS}\\[0.2cm]
{E-mail: grushevskaja@bsu.by}
\end{center}
 \parindent 24pt

\begin{quotation}

{ \bf \centerline{Abstract}  }

Variation principle has been developed to calculate many-particle effects
in crystals.
 Within the framework of quasi-particle concept   the variation
principle has been used to find one-electron states with taking
into account of  effects due to  non-locality of electronic
density functional in electromagnetic fields. A secondary quantized  density
matrix was used to find  the Green function of a quasiparticle and changes of its
effective mass due to correlated motion of interacting electrons.
\end{quotation}

\section{
Introduction}


Because of high dimensionality the complexity of many-electron
problems  all problems of a solid state physics of this type are
much higher  than of one-electron problem. Results of a
solution of the one-electron problem can be utilized  in many-electron
problems in solid state physics if one supposes that  motion of
an electron happens in a self-consistent one-particle potential $V
(\vec r) $. The self-consistent potential is yielded by the solution of the
 Poisson
equation, for which the density of one-electron states  is a
self-consistent solution of the one-particle Hartree - Fock
equations  \cite{Hartry} - \cite{Froese}. These equations have
been written originally to calculate one-particle states of
many-electron atom. Equations of the Hartree - Fock type for band
calculations have been proposed by Kohn and Sham \cite{Kon_SHam}.
It is practically impossible to look for  their solution  without
additional assumptions if many-particle effects are circumscribed by
distributed in space  electron density functional. One of them is that one-particle solutions
describe a crystal as a set of  interacting quasiparticle
excitations \cite {Ary}. In a given paper we offer an approach which
allows to determine one-electron states with taking into account
of  the effects owing to a non-locality of  electron density
functional in electromagnetic fields and which is based on usage of
the variation principle.
%
%

\section{
A variational principle for band calculations}

%
Let us write a one-electron Hartree - Fock Hamiltonian for a
system of $N$ electrons, $N\to \infty $, moving in a field originated by atoms
nucleus' of a crystal
\begin{eqnarray}
\left[{1\over 2} \hat p^2_i + U(r_i)
 +  \hat V^{sc} (\vec r_i, \sigma_i) - \hat \Sigma ^x(\vec r_i, \sigma_i)
 \right] \psi_n (k_i r_i)=
 \nonumber \\
 =(\epsilon_n(0)+ \epsilon_n(k_i)) \psi_n (k_i r_i)
 \label{hartry-fock-eqs}
\end{eqnarray}
with a unit choice $\hbar =1, \ m=1$,
where $\hat p^2_i /2 = -{1\over 2} \triangle (\vec r_i)$ is a kinetic energy
in system of atomic units,
$\triangle (\vec r_i)$ is the Laplacian operator written in a given point
with a radius-vector
$\vec r_i$ in which
$i$-th  electron 
having a spin $\sigma_i$ is situated; $U(r_i)$
is  potential energy of $i$-th 
electron in nucleus field of the crystal, $\hat V^{sc}$ and 
$ \hat \Sigma ^x $ are
operators of Coulomb and exchange interactions, respectively \cite{Vesel-Labzov}:
\begin{eqnarray}
\hat V^{sc}(\vec r_i, \sigma_i)\psi_n (k_i r_i)=\sum_{m=1}^N
\int \psi_m^* (k_i r'_i) v(|\vec r_i - \vec r'_i|)\psi_m (k_i r'_i)
\ d   r'_i \psi_n (k_i r_i)
,\nonumber \\
\label{coulon}\\
\hat \Sigma ^x(\vec r_i, \sigma_i) \psi_n (k_i r_i)
=\sum_{m=1}^N
\int \psi_m^* (k_i r'_i) v(|\vec r_i - \vec r'_i|)\psi_n (k_i r'_i)
\ d  r'_i \psi_m (k_i r_i); \nonumber \\
\label{exchange}
\end{eqnarray}
$r_i\equiv \{\vec{r}_i, \sigma_i\}$, $\psi_n (k_i r_i)$ is a
wave function including spin and coordinate parts,
$v(|\vec r_i - \vec r'_i|)$ is  potential energy of electron interaction.
%
The physical sense of the operators (\ref{coulon}, \ref{exchange})
becomes obvious if one rewrites them in terms of spin-zero electronic
density $ \rho (\vec r, \vec r ') $ and assumes that
the interaction $v $ is the Coulomb one:
\begin{eqnarray}
\rho(\vec r, \vec r')={1\over 2}\sum_{m=1}^{N-1}\left(
 \psi_m^* (\vec k  \cdot \vec r ,\sigma )
\psi_m (\vec k  \cdot \vec r' , -\sigma )+
\psi_m^* (\vec k  \cdot \vec r ,-\sigma)
\psi_m (\vec k  \cdot \vec r', \sigma )
\right)
\nonumber\\
=\sum_{m=1}^{(N-1)/2}
\psi_m^* (\vec k  \cdot \vec r)
\psi_m (\vec k  \cdot \vec r'),\quad v =e^2/|\vec r - \vec r'|.
\end{eqnarray}
%
It follows from here that the operator $ \hat V ^ {sc} $ gives
electrostatic interaction of one electron with an electron density
created by residual $N-1 $  electrons, with an electrostatic
self-action (s.a.):
\begin{eqnarray}
\hat V^{sc}(\vec r_i, \sigma_i)\psi_n (k_i r_i)=
\sum_\sigma \sum_{m=1}^{(N-1)/2}
\int \psi_m^* (\vec k_i \cdot \vec r'_i) v(|\vec r_i - \vec r'_i|)
\psi_m (\vec k_i \cdot \vec r'_i)
\ d   \vec r'_i \psi_n (\vec k_i \cdot \vec r_i)
+\mbox{s.a.}
\nonumber \\
=2 \int \ d   \vec r'_i
{e^2 \rho(\vec r'_i, \vec r'_i)\over |\vec r_i - \vec r'_i|}
\psi_n (k_i r_i)
+\mbox{s.a.}
\end{eqnarray}
%
Analogously we get that the operator $ \hat \Sigma ^x $ gives a
quantum exchange with an exchange self-action (s.a.):
\begin{eqnarray}
\hat \Sigma ^x(\vec r_i, \sigma_i) \psi_n (k_i r_i)
=\sum_{m=1}^{N-1}
\int \int d  r_j d \sigma_j \psi_m^* (\vec k_i \cdot \vec r_j, \sigma_j)
v(|\vec r_i - \vec r_j|)\psi_n (k_i\cdot \vec r_j, \sigma_j)
 \psi_m (\vec k_i \cdot \vec r_i, \sigma_i) \nonumber \\
 ={1\over 2}\sum_{m=1}^{N-1}\int \int d  r_j d \sigma_j
 \left(
 \psi_m^* (\vec k  \cdot \vec r_j ,\sigma_j )
\psi_m (\vec k  \cdot \vec r_i , -\sigma _j) \delta(\sigma_j - \sigma_i )\right.\nonumber \\
\left. +
\psi_m^* (\vec k  \cdot \vec r_i ,-\sigma_j)
\psi_m (\vec k  \cdot \vec r_j, \sigma_j)\delta(\sigma_j - \sigma_i )
\right)
v(|\vec r_i - \vec r_j|)\psi_n (k_i\cdot \vec r_j, \sigma_j)+\mbox{s.a.}
\nonumber \\
 =\int \ d   \vec r_j
{e^2 \rho(\vec r _j, \vec r _i)\over |\vec r_i - \vec r _j|}
\psi_n (k_i r_j)
+\mbox{s.a.}
\label{exchange1}
\end{eqnarray}
%
Since operators $ \hat V ^ {sc} $ and $ \hat \Sigma ^x $ in
expression (\ref{hartry-fock-eqs}) are subtracted one from
another the self-acting terms vanish.


A quantity $ \epsilon_n (k_i) $ entering in expansion $E_n (k_i)
=f (\epsilon_n (0)) + \epsilon_n (k_i) $ being $n $-th eigenvalue
$E_n (k_i) $ of the Hamiltonian for $N $ particles system in
Hartree -Fock approximation is the energy of $n $-th band. A
quantity $ \epsilon_n (0) $ determines a reference point $ \mbox
{Extr} E_n (k_i) =f (\epsilon_n (0)) $ of  $n $-th bands, where $f
$ is an unknown function. Within the framework of quasiparticle
concept the physical sense of $ \epsilon_n (k_i) $ is the energy of
quasiparticle excitation.


Let us take a solution of  one-electron problem for an atomic
area in a cell potentials approximation as $ \epsilon_n (0) $ and
a basic set on which one constructs an expansion of a trial
function $ \widetilde \psi_n (k_i r_i) $ of $n $-th band. Then one
can realize  a variational principle as the following expression:
\begin{eqnarray}
\delta \epsilon_n[\widetilde \psi_n] =0 ,
\label{exitation-variation-princip}
\end{eqnarray}
which is  determined not strictly as it is used to find excited
states for which, as a rule,  the basic set of functions  is
unknown \cite{Vesel-Labzov}.
%

%
Let us show, that the variational principle for excited
states in the form (\ref{exitation-variation-princip}) can be
made strictly determined one.
 Since, by definition, the energy of a nonexcited state coincides with an extremum of a functional $E_n [\widetilde \psi_n] $ we have the following equality of variations:
\begin{eqnarray}
 \delta \epsilon_n[\widetilde \psi_n]= \delta E_n[\widetilde \psi_n]  =0 .
\label{variation-princip}
\end{eqnarray}
%
The ambiguity of a variation (\ref{variation-princip}) consists
only in an arbitrariness of a position of a reference point for
an energy band as the variation procedure for an one-electron
problem is strictly determined. If a symmetry of atomic areas is
definite and a cell partition of the  crystalline space is unambiguous
then, in principle, this arbitrariness is removed easy by
utilizing the quasi-particles concept, according to which
\begin{eqnarray}
\epsilon_n(k_i) = \epsilon_n(-k_i).
\label{variation-princip1}
\end{eqnarray}
%
It  follows  from the expressions (\ref {variation-princip}) and
(\ref {variation-princip1}), that the additional variation of a
reference point of a band gives a coincidence of this point with a
centre of the energy band for a crystal and allows to transform
the variational principle (\ref {variation-princip})  finding
noninteracting quasiparticles  states  to strictly determined
expression
\begin{eqnarray}
\delta E_n[\widetilde \psi_n] = 0,
\quad   \delta f[\widetilde \psi_n]=0.
\label{variation-princip2}
\end{eqnarray}
Let us find the reference point $ \mbox {Extr} E_n (k_i) $ of
energy band using a method of density matrix functional.
%

\section{Secondary quantized reduced density matrix
}

%
As is known \cite{faddeev}, equations in quantum mechanics can be
written not for a wave function, but for a density matrix $ \rho
$. The operator $ \rho $ is a projective operator for pure states
and can be presented in terms  of  Dirac ket(bra)-vectors as $
\rho = | \psi \rangle \langle \psi | $. If the operator $ \rho $
is known then we can find, by definition,  energy $E $ of system
of $N $ particles, described by a Hamiltonian $ \hat H (\vec
r_1, \ldots, \vec r_N) $ and a wave function $ | \psi \rangle $ as
\begin{eqnarray}
E= \mbox{Sp }\rho \hat H . \label{mean-energy}
\end{eqnarray}

%
Let us introduce  $n$-dimensional density matrix $\rho_n$ by
\begin{eqnarray}
\rho _n(\vec r_1',\ldots, \vec r_n'; \vec r_1,\ldots, \vec r_n) =
{N!\over (N-n)!}
\int \ d\vec r_{n+1}\ldots \ d\vec r_N
\nonumber \\
\times \langle
\vec r_1',\ldots, \vec r_n', \vec r_{n+1},\ldots,\vec r_N|\psi \rangle
\langle \psi |\vec r_1,\ldots, \vec r_n , \vec r_{n+1},\ldots, \vec r_N\rangle
,
\label{n-particle-function}
\end{eqnarray}
%
which, by implication, is a  reduced coordinate distribution
$n$-particle function \cite{Lovdin}. By definition, it has a
normalization
\begin{eqnarray}
\mbox{Sp} \rho_n ={ N!\over (N-n)!} \quad n=1,\ldots ,N .
\end{eqnarray}

Let us examine the system consisted of $N$ with a pairwise interaction
$$\sum _{i>j=1}^N v(|\vec r_i - \vec r'_i|).$$
Then, since projective operators possess the following properties:
$ \rho^2= \rho$ and 
$ \rho^*= \rho$  the expression  
(\ref{mean-energy}) is transformed to the form: 
\begin{eqnarray}
E=\mbox{Sp }\rho \left(\hat H_0 +\sum _{i>j=1}^N v(|\vec r_i - \vec r_j|)\right)
=\mbox{Sp }\rho \sum _{i=1}^N\hat h(\vec r_i) 
+
\mbox{Sp }\rho ^2\sum _{i>j=1}^N v(\vec r_i,  \vec r_j)=
\nonumber \\
=\mbox{Sp }\rho \sum _{i=1}^N\hat h(\vec r_i) 
+
\mbox{Sp }|\rho |^2\sum _{i>j=1}^N v(\vec r_i,  \vec r_j)
, \label{mean-energy1-0}
\end{eqnarray}
where
$\hat h(\vec r_i)={1\over 2} \hat p^2_i + U(r_i)$.
Since the non-perturbed hamiltonian 
$\hat H_0$ in Eq.~
(\ref{mean-energy1-0}) consists of independent one-particle summands
and  interaction of particles occurs pairwise,
the operator of trace appears in Eq.~(\ref{mean-energy1-0}) over one
variable $r_1$ and two variables 
$r_1,r_2$:
\begin{eqnarray}
E=\mbox{Sp }\sum_{i=1}^N \hat h(\vec r_1)
\int \ d\vec r_{2}\ldots \ d\vec r_N
\nonumber \\
\times \langle
\vec r_1',\vec r_{2},\ldots,\vec r_N|\psi \rangle
\langle \psi |\vec r_1,\ldots, \vec r_N\rangle
 +
\mbox{Sp } \sum_{i>j=1}^N v(|\vec r_1 - \vec r_2|)
\nonumber \\
\times {1\over 2}\left[
\int \ d\vec r_{1}d\vec r_{3}\ldots \ d\vec r_N \left|\langle
\vec r_1,\vec r'_{2},\vec r_3, \ldots,\vec r_N|\psi \rangle
\langle \psi |\vec r_1,\ldots, \vec r_N\rangle \right|^2
\right.
\nonumber \\
+ \left.
\int \ d\vec r_{2}\ldots \ d\vec r_N \left|\langle
\vec r_1',\vec r_{2},\vec r_3, \ldots,\vec r_N|\psi
\rangle
\langle \psi |\vec r_1,\ldots, \vec r_N\rangle \right| ^2
\right]
, \label{mean-energy1}
\end{eqnarray}
Using the equality:
$\int \vec k \cdot \vec k d\theta =2 |k|^2$ one can transform
Eq.~(\ref{mean-energy1}) and obtain: 
\begin{eqnarray}
E=
\mbox{Sp }\hat h(\vec r_1)
N\int \ d\vec r_{2}\ldots \ d\vec r_N
\nonumber \\
\times \langle
\vec r_1',\vec r_{2},\ldots,\vec r_N|\psi \rangle
\langle \psi |\vec r_1,\ldots, \vec r_N\rangle
 +{1\over 4}\hspace{4mm}
\mbox{Sp } v(|\vec r_1 - \vec r_2|)N(N-1)
\nonumber \\
\times \left[
\int \ d\vec r_{1}d\vec r_{3}\ldots \ d\vec r_N \langle
\vec r_1,\vec r'_{2},\vec r_3, \ldots,\vec r_N|\psi \rangle
\right.
\times \nonumber \\
\times
\int \ d\vec r_{2}\ldots \ d\vec r_N \langle \psi
|\vec r_1',\vec r_{2},\vec r_3, \ldots,\vec r_N
\rangle
\langle \vec r_1,\ldots, \vec r_N|\psi \rangle
\langle \psi |\vec r_1,\ldots, \vec r_N\rangle
\nonumber \\
+
\int \ d\vec r_{2}\ldots \ d\vec r_N \langle
\vec r_1',\vec r_{2},\vec r_3, \ldots,\vec r_N|\psi
\rangle \times
\nonumber \\
\left.
\times \int \ d\vec r_{1}d\vec r_{3}\ldots \ d\vec r_N \langle
\psi | \vec r_1,\vec r'_{2},\vec r_3, \ldots,\vec r_N \rangle
\langle \vec r_1,\ldots, \vec r_N|\psi \rangle
\langle \psi |\vec r_1,\ldots, \vec r_N\rangle
\right] . \nonumber \\
\label{mean-energy2}
\end{eqnarray}
%
It is easy to see, that the first summand from the right-hand side of
Eq.~(\ref {mean-energy2}) contains the reduced one-particle
density matrix $ \rho_1 $ as a multiplier. Therefore after some
obvious transformations Eq.~(\ref {mean-energy2}) can be rewritten
as:
\begin{eqnarray}
E=
\mbox{Sp }\hat h(\vec r_1)
\rho_1(\vec r_1',\vec r_1)
 +{1\over 2}
\mbox{Sp } v(|\vec r_1 - \vec r_2|)N(N-1)
\nonumber \\
\times
\int \ d\vec r_{3}\ldots \ d\vec r_N \langle
\vec r'_1,\vec r'_{2},\vec r_3, \ldots,\vec r_N|\psi \rangle
\langle \psi
|\vec r_1,\vec r_{2},\vec r_3, \ldots,\vec r_N
\rangle
\times \nonumber \\
\times
\int \ d\vec r_{1} d\vec r_{2}\ldots \ d\vec r_N
\langle \vec r_1,\ldots, \vec r_N|\psi \rangle
\langle \psi |\vec r_1,\ldots, \vec r_N\rangle
\nonumber \\
=
\mbox{Sp }\hat h(\vec r_1)
\rho_1(\vec r_1',\vec r_1)
 +{1\over 2}\mbox{Sp } v(|\vec r_1 - \vec r_2|)N(N-1)
\nonumber \\
\times
\int \ d\vec r_{3}\ldots \ d\vec r_N \langle
\vec r'_1,\vec r'_{2},\vec r_3, \ldots,\vec r_N|\psi \rangle
\langle \psi
|\vec r_1,\vec r_{2},\vec r_3, \ldots,\vec r_N
\rangle .
\nonumber \\
\label{mean-energy3}
\end{eqnarray}
Here one takes into account the normalization of wave function
$|\psi \rangle $: $\int \langle \psi |\psi \rangle  \ d\vec r_{1}
d\vec r_{2}\ldots \ d\vec r_N=1$.

%
If we observe that in the right side of the equation (\ref
{mean-energy3}) the second summand contains the reduced
two-particle density matrix $ \rho_2 $ as a multiplier then it is
possible to transform this equation to the expression which has
the reduced  density matrixes $ \rho_1 $ and $ \rho_2 $:
\begin{eqnarray}
E=
\mbox{Sp }\hat h(\vec r_1)
\rho_1(\vec r_1',\vec r_1)
 +{1\over 2}\mbox{Sp }
  v(|\vec r_1 - \vec r_2|)
  \rho_2(\vec r'_1,\vec r'_{2};\vec r_{1}, \vec r_{2})= \nonumber \\
=
\epsilon^{(0)} N +{1\over 2}\mbox{Sp }
  v(|\vec r_1 - \vec r_2|)
  \rho_2(\vec r'_1,\vec r'_{2};\vec r_{1}, \vec r_{2}) \label{mean-energy4},
\end{eqnarray}
where $\epsilon^{(0)}$ is the one-electron state,
the summand $ 
E^{HF}={1\over 2}\mbox{Sp }
  v(|\vec r_1 - \vec r_2|)$
  \noindent $
  \rho_2(\vec r'_1,\vec r'_{2};\vec r_{1}, \vec r_{2})
$ is excitation energy of the system under the interaction.
Since in the Hartree - Fock approximation the reduced density
matrix $ \rho_2 $  is factorized, at first, the
energy $\epsilon ^ {(0)} N $ equal to energy of  $N $ one-electron
states, including  kinetic energy of an electron, energy of an
electron in a self-consistent scalar potential, and  exchange
energy, and, secondly,  energy $E ^ {HF} $ of the excitation
yield,  as it follows from the equations
(\ref {hartry-fock-eqs}) and (\ref {mean-energy4}), a contribution to the
electronic energy $E $ of a crystal.
Further we shall show, that the Hartree - Fock approximation is an
one-particle approximation in the sense that in this approximation
the excitation energy $E ^ {HF} $ is represented as the energy of
quasiparticle states.
%

Now, we rewrite 
Eq.~(\ref{hartry-fock-eqs}) in the representation of Dirac ket(bra)-  vectors:
\begin{eqnarray}
\hat h(k) | n; k \rangle + \sum_{m=1, m\neq n}^N
\int \delta(k-k') \ dk' (| n; k \rangle \langle  m; k' | v(kk')| m; k'\rangle
+\nonumber \\
+ (| n; k' \rangle \delta_{nm})\langle  m; k' | v(kk')
(| m; k\rangle \delta_{mn} ))
\nonumber \\
-
\sum_{m=1}^N
\int | m; k \rangle \langle  m; k' | v(kk')| n; k'\rangle  \delta(k-k') \ dk'
=
\nonumber \\
= | n; k \rangle (\epsilon_n(0)+ \epsilon_n(k))
\label{moment-Hartry-Fock-eq}
\end{eqnarray}
where
$k_i\equiv \{\vec{k}_i, \sigma_i\}$, $\hat h(k)$ is
a momentum representation of the non-perturbed hamiltonian,
\noindent
$v(kk')=\int d \vec r d\vec r'
|\vec r\rangle \langle  \vec k\cdot \vec r |
v(|\vec r - \vec r'|)
| \vec k'\cdot \vec r '\rangle \langle \vec r' |$
is a momentum representation of the Coulomb interaction operator,
$\delta(k-k')$ is the Dirac 
$\delta $-function manifesting the presence of the law of conservation of
momentum.

Let us introduce projective operators $\hat\rho ^{mn}_{kk'}$
\begin{eqnarray}
\hat \rho ^{mn}_{kk'} \equiv | m; k'\rangle \langle  n; k|
\label{secondary-quant-projector}
\end{eqnarray}
and express 
Eq.~(\ref{moment-Hartry-Fock-eq}) via these operators 
$\hat \rho ^{mn}_{kk'}$.
%
For this purpose, Eq.~(\ref {moment-Hartry-Fock-eq}) is multiplied
  on the right by bra-vector $ \langle n; k | $. Then, additional summating over
$n $ and integrating over $dk $  one get the  equation:
\begin{eqnarray}
\sum_{n=1}^N \int dk\ \hat h(k) | n; k \rangle \langle  n; k|+ \sum_{n=1}^N
\int \int \delta(k-k')\ dk\ dk'
\times \nonumber \\
\times
| n; k \rangle \sum_{m=1}^N \langle  m; k' | v(k,k')| m; k'\rangle
\langle  n; k|
+ \int \int \delta(k-k')\ dk\ dk' \times \nonumber \\
\times
\left(\sum_{m=1}^N| n; k '\rangle \delta_{nm}\langle  m; k' |\right) v(kk')
\left(\sum_{ n=1}^N| m; k\rangle \delta_{mn}\langle  n; k| \right)
-\nonumber \\
-
\int \int \sum_{m=1}^N | m; k \rangle \langle  m; k' | v(kk')
\sum_{n=1}^N| n; k'\rangle \langle  n; k|
\delta(k'-k) d(-k) d(-k')
\nonumber \\
= \int dk \sum_{n=1}^N \langle  n; k|n; k \rangle (\epsilon_n(0)
+ \epsilon_n(k)).
\nonumber \\
\label{moment-Hartry-Fock-eq1}
\end{eqnarray}
%
We see that the first and second terms on the left of the equation
(\ref {moment-Hartry-Fock-eq1}) are traces of a matrix
representation of operators $ \hat \rho \hat h $ and $ \hat \rho
^* \hat \rho v $, and the third and fourth terms on the left of
equation (\ref {moment-Hartry-Fock-eq1}) are mutually cancelled.
Hence, it means that using  normability of function $
|n; k \rangle $: $ \int dk\langle n; k|n; k \rangle =1 $, we
obtain the following equation:
\begin{eqnarray}
\mbox{Sp}   \hat \rho \hat h
+ \mbox{Sp} \hat \rho ^* \hat \rho v
=\epsilon_n(0)N +\int dk \sum_{n=1}^N \langle  n; k|n; k \rangle
\epsilon_n(k).
\label{moment-Hartry-Fock-eq2}
\end{eqnarray}
Using properties of the projective operators $\hat\rho ^{mn}_{kk'}$:
$\left(\hat\rho ^{mn}_{kk'}\right)^*=\hat\rho ^{mn}_{kk'}$ and 
$\left(\hat\rho ^{mn}_{kk'}\right)^2=\hat\rho ^{mn}_{kk'}$
one can transform 
Eq.~(\ref{moment-Hartry-Fock-eq2}) to the form: 
\begin{eqnarray}
\mbox{Sp}   \hat \rho (\hat h +  v)
=\epsilon_n(0)N +\int dk \sum_{n=1}^N \langle  n; k|n; k \rangle
 \epsilon_n(k)= \epsilon_n(0)N + \epsilon.
\label{moment-Hartry-Fock-eq3}
\end{eqnarray}

%
%
Let us elucidate a physical sense of introduced projective
operators $ \hat\rho ^ {mn} _ {kk '} $. It follows from comparison
(\ref {mean-energy4}) and (\ref {moment-Hartry-Fock-eq3}) that  the
energy of a quasi-particle $ \epsilon $ is  on the right of
Eq.~(\ref {moment-Hartry-Fock-eq3}) accurate to the constant $
\epsilon_n {(0)} N $. It follows from here that the operator $
\hat\rho ^ {mn} _ {kk '} $ allows to calculate  energy $
\epsilon $ of quasiparticle excitations. It means that the
expression (\ref {moment-Hartry-Fock-eq3}) is  nothing else but
a procedure of average on density matrix. Since the averaging with
the help of the operator $ \hat\rho ^ {mn} _ {kk '} $ yields
energy $ \epsilon $ of quasi-particle this operator is a secondary
quantized density matrix.

It  follows from the comparison of right sides of the equations (\ref
{hartry-fock-eqs}), (\ref {mean-energy4}), and (\ref
{moment-Hartry-Fock-eq3}) that the reference point $ \mbox {Extr}
E_n (k_i) $ of energy band determines the solution $ \epsilon_n
{(0)} $ of the one-electron problem
\begin{eqnarray}
\epsilon_n{(0)} =\mbox{Extr } E_n(k_i)/N  \label{reference-point}.
\end{eqnarray}

Thus, one has proved that the equation (\ref {hartry-fock-eqs})
can be considered as an equation describing a state of
quasi-particle and determining its energy accurate to the constant
$ \epsilon_n {(0)} N $.

\section{Green function for one-particle state
}
%
It follows from Eq.~(\ref {moment-Hartry-Fock-eq3}) also  that the
quantity $ \epsilon_n {(k_i)} $ can be interpreted as an
eigenvalue of the Hamiltonian for the quasi-particle excitation
without taking into account  interaction of quasi-particles.
Therefore, the equation (\ref {moment-Hartry-Fock-eq3}), written
in the formalism of  density matrix $ \hat\rho ^ {(0)} _ {nn '; kk
'} \equiv \hat\rho ^ {mn} _ {kk '} $ can be rewritten in the
formalism of wave functions in coordinate representation and in a
limit of large $N $, $N\to \infty $ in the following way:
\begin{eqnarray}
 \left(\imath {\partial \over \partial t }-(\hat h +  \Sigma^{x}+V^{sc} )\right)
 \sum_n   \hat\rho ^{(0)}_{nn'; rr'}
=\lim_{N\to\infty}  (-\epsilon_n(0))N \delta_{rr'},
\label{moment-Hartry-Fock-eq4}
\end{eqnarray}
where property $ \int dk\langle  n; kr|n; kr'
\rangle =\delta_{rr'}$ has been used; $\delta_{rr'}$ is a delta symbol.
Since the energy $ \epsilon_n (0) $ of  the bound one-electron
state is negative: $ \epsilon_n (0) <0 $, the right side of Eq.~
(\ref {moment-Hartry-Fock-eq4}) represents   itself a Dirac
$\delta$-function $ \delta(r-r')$. It allows to write Eq.~(\ref
{moment-Hartry-Fock-eq4}) as:
\begin{eqnarray}
 \left(\imath {\partial \over \partial t }-\hat h^{HF} \right)
 \sum_n   \hat\rho ^{(0)}_{nn'; r,r'}
=\delta(r-r'),
\label{moment-Hartry-Fock-eq5}
\end{eqnarray}
where  
$\hat
h^{HF}=(\hat h + \overline\Sigma^{x}+V^{sc})$, $\overline
\Sigma^{x}=-\Sigma^{x}$. 
Eq.~(\ref{moment-Hartry-Fock-eq5}) is the equation for the Green
function. 
It means that in the secondary quantized representation an operator
\begin{eqnarray}
\hat G_1^{(0)}(n'; r,r') = \sum_n \hat\rho ^{(0)}_{nn'; r,r'}
\end{eqnarray}
 possesses properties of non-perturbed Green function.

%
So, the quasi-particle excitation determined by the Hamiltonian $
\hat h ^ {HF} $ can be considered as a free particle whose
equation of motion is the equation (\ref{moment-Hartry-Fock-eq5}).

%
In the many-body problem, in particular, in calculations of an
energy-band crystal structure   a contribution given by interaction of
electromagnetic field with matter is played the essential role. To take into
account many-particle effects due to a
correlated motion of electron we should describe the system
by self-consistent solutions of the non-stationary equation
\begin{eqnarray}
 \imath {\partial\Psi (t) \over \partial t }= \hat H \Psi (t)
\label{non-stationary-quant-motion-eq}
\end{eqnarray}
where $\hat H$ is a Schr\"odinger hamiltonian in a
non-relativistic case or a Dirac hamiltonian in a relativistic
case.
%
It turns  the variational principle (\ref {variation-princip2})
into a variational principle for the excited states, not being
strictly definite one. Further, we shall show that within the
framework of the concept of quasiparticle excitations this
ambiguity can be removed by means of the account of interaction
as, at first,  a change of a quasi-particle mass and, second, as
a changing of the location of reference point of energy band.

%
We have proved that for the secondary quantized representation the
operator $ \rho $ looks as $ \hat \rho = | \hat \psi \rangle
\langle \hat\psi | $ and possesses properties of the Green
function $G_1 $. Therefore the sum $ \hat G_1 $ over $n $ from
elements of the matrix $ \hat\rho ^ {nn '} _ {kk '} $ for the
secondary quantized density matrix $ \hat \rho $ describing an
interacting particle satisfies a Dyson equation in a
nonrelativistic case or to a  Schwinger - Dyson equation in a
relativistic case:
\begin{eqnarray}
G_1(1;2)= G_1^{(0)}(1;2)+\int d3\ d4\ G_1^{(0)}(1;3) \hat \Sigma (3,4) G_1(4;2)
\label{Shwinger-Dayson-eq}
\end{eqnarray}
$G_1^{(0)}(1;2)$ is a free Green function,
$\hat \Sigma (3,4)$ is a self-energy operator:
$\hat \Sigma = {\overline \Sigma^{x}}+\hat \Sigma^{c}$, $\hat
\Sigma^{c}$ is a correlation interactions,
representing itself a part of the self-energy which describes the
many-particle effects.
Here numerical labels for the arguments are used: $ \{r_1, t_1 \}
= x_1\equiv 1 $, etc. Acting on Eq.~(\ref {Shwinger-Dayson-eq}) by
the operator $ \imath {\partial \over \partial t}-\hat h ^ {HF} $
and using the equation of motion for the free particle (\ref
{moment-Hartry-Fock-eq5}) we get the equation for the perturbed
Green function as
\begin{eqnarray}
\left[ \imath {\partial \over \partial t }-\hat h^{HF}(r_1)\right]G_1(n';1,2)-
\int d3 \hat \Sigma^{c} (n';1,3) G_1(n';3,2)= \nonumber \\
=
(-\epsilon_n(0))N \delta_{r_1r_2}.
\label{Perturbed-Green-func-eq}
\end{eqnarray}
Rewriting Eq.~(\ref{Perturbed-Green-func-eq}) in the formalism of
wave functions one gets
\begin{eqnarray}
\left[\imath {\partial \over \partial t }-\hat h^{HF}(r_1)\right]\psi_{n}
(k_1r_1) -
\int d \vec r_2 \hat \Sigma^{c} (n;1,2) \psi_{n}(k_1r_2)=
\nonumber \\
(-\epsilon_n(0)) \psi_{n}
(k_1r_1).
\label{Perturbed-wave-func-eq1}
\end{eqnarray}
Since the expression:  
$\imath {\partial \psi_{n}\over \partial t} =\epsilon_n(k_1)$
takes place, then
Eq.~(\ref{Perturbed-wave-func-eq1}) yields  
the Hartree - Fock taking into account of interacting
quasi-particles
\begin{eqnarray}
 \hat h^{HF}(r_1)\psi_{n}
(k_1r_1) + \int d \vec r_2 \hat \Sigma^c (n;1,2) \psi_{n}(k_1r_2)
=\nonumber \\
=(\epsilon_n(0)+\epsilon_n(k_1)) \psi_{n}
(k_1r_1).
\label{Perturbed-Hartry-Fock-eqs}
\end{eqnarray}

Let us define a mass operator $\widehat {\Delta M} $ as: 
\begin{eqnarray}
\widehat {\Delta M}\psi_n (k_1 r_1)
 = \int d \vec r_2 \hat \Sigma^c (n;1,2) \psi_{n}(k_1r_2).
\end{eqnarray}

Within the framework of the concept quasiparticle excitations it
is possible to represent the operator $ \widehat {\Delta M} $ in
the diagonal form:
\begin{eqnarray}
\widehat {\Delta M}\psi_n (k_i r_i)
 =(\Delta M_n(0)+ \Delta M_n(k_i)) \psi_n (k_i r_i).
\end{eqnarray}
It is known that the eigenvalue of mass operator possesses the
property: $\Delta M_n(k_i)=\Delta M_n(-k_i)$. Here 
$\Delta M_n(0)$ is an eigenvalue of mass operator 
$\widehat {\Delta M}$ in the limit 
$\vec k \to 0$.

It follows from here the physical sense of $ \widehat {\Delta M} $.
It determines an effective mass of the quasi-particle and an
efficient reference point of the energy band:
\begin{eqnarray}
 \hat h^{HF}(r_1)\psi_{n}(k_1r_1) =
(\tilde \epsilon_n(0)+\tilde \epsilon_n(k_1))\psi_{n}(k_1r_1)\equiv \nonumber \\
 \equiv
\left[(\epsilon_n(0)+\Delta M_n(0))+(\epsilon_n(k_1)+\Delta M_n(k_1))\right]
 \psi_{n}(k_1r_1).
\label{Perturbed-Hartry-Fock-eqs1}
\end{eqnarray}

%
Since the change of the mass of quasi-particle determined by the
operator $ \widehat{\Delta M}$ maintains the condition (\ref{variation-princip1})
then if to take into account the change of
the reference point of band at interaction, the variational
principle for the interacting system becomes a strictly definite one
and takes the form:
\begin{eqnarray}
\delta E_n[\widetilde \psi_n] = 0,
\quad   \epsilon_n(0)= {1\over N}\mbox{Extr} E_n(k_i)- \Delta M_n(0).
\label{variation-princip3}
\end{eqnarray}

\noindent\underline{Equation of motion for one-particle state and basis set
of wave functions for exited atom
}

Let us consider a Green function normalized per unit volume $V=1 $
so that an  average energy in $V $ is equal to the energy of the
one-particle state and $N=1 $. If $-\epsilon_n (0) \to \infty $
then Eq.~(\ref {Perturbed-Green-func-eq}) describes a propagation
of one particle and should be rewritten as
\begin{eqnarray}
\left[ \imath {\partial \over \partial t }-\hat h^{HF}(r_1)\right]G_1(n';1,2)-
\int d3 \hat \Sigma^c (n';1,3) G_1(n';3,2)= \nonumber \\
=
(-\epsilon_n(0)) \delta_{r_1r_2}.
\label{Particle-Green-func-eq}
\end{eqnarray}
%
It follows from here that according to the definition of Green
functions we have the following expression for the energy $
\epsilon_n (0) $:
\begin{eqnarray}
 -\epsilon_n(0)=C - a_n, \quad C \to \infty; \label{Particle-energy}
\end{eqnarray}
where $a_n$ is a finite quantity. 
Hence, since the energy is counted off from an arbitrary value,
Eq.~ (\ref {variation-princip3}) yields the following expression for reference
points $
\epsilon (0) ^ {\pm}_n $ of a quasi-particle energy and a
antiquasi-particle energy
\begin{eqnarray}
  \epsilon(0)^{\pm}_n\equiv \pm a_n
  =  \left( \mbox{Extr} \tilde{E}(k_1)\mp \Delta {M}_n(0) \right )/2.
\label{zone-reference}
\end{eqnarray}
Here one took into account that
$N=1$;  
an extremum of zone is redefined as
$\mbox{Extr}\tilde{E}(k_1)= \mbox{Extr}{E}_n(k_1)-C_n$,
the sign $\{\pm\}$ in left-hand side denotes a case of quasiparticles and
antiquasi\-par\-tic\-les, respectively;
and the energy of particles in the pair is counted off from zero level.
One gets from the expression (\ref{zone-reference}) 
that $a_n$ is the energy which is required to create a pair from
quasiparticle  and antiquasiparticle
when $k_1=0$ because  
\begin{eqnarray}
a_n=(\epsilon(0)^{+} _n- \epsilon(0)^{-}_n)/2. \label{gap}
\end{eqnarray}


Because of an additional term $ \tilde \epsilon_n (0) $ in a
right-hand side of  Eq. (\ref {Perturbed-Hartry-Fock-eqs1}) we,
generally speaking, cannot examine the left-hand side as a
Hamiltonian operator of the quasi particles system acting on a
corresponding wave function and as a consequence,  can not
construct a basis set of one-particle states of the problem. However,
further we show, that $ \hat h ^ {HF} $ is a Hamiltonian of an
electron - hole pair.

{\it 
Non-relativistic case }

One can examine in non-relativistic limit quantum systems which
are characterized by a small value of  $\Delta {M}_n(0)$:
\begin{equation}
\Delta {M}_n (0)\to 0. \label{light-electron}
\end{equation}
It means that weak many-particle effects occur and, accordingly,
we can speak about a "light" \  electron. The equality (\ref
{zone-reference})  occurs under condition of
(\ref{light-electron})only in the case if $a_n=0 $. From here  it
follows, that the energy $a_n $ of a pair is equal to zero. In other
words,  the energy is not expended to create an electron - hole
pair.

Substituting Eqs.~(\ref{Particle-energy}), (\ref{light-electron})
into 
Eq.~(\ref{Perturbed-Hartry-Fock-eqs1}) and taking into account the
condition $a_n=0$, one gets the 
 Schr\"odinger equation as
\begin{eqnarray}
 \hat h^{HF}(r_1)\psi_{n}(r_1) =
\tilde {\tilde \epsilon}_n \psi_{n}(r_1),
\label{Perturbed-Hartry-Fock-eqs2}
\end{eqnarray}
which describes the quasiparticle - antiquasiparticle pair (a
non-relativistic electron - hole pair). Here $\tilde {\tilde
\epsilon}_n={\tilde \epsilon}_n -C$.

Since the  energy $a_n $, expended on  creation of a pair, equals
to zero we have proved that the variable $ \tilde {\tilde
\epsilon} _n $ can be understood as the energy of an electron -
hole pair. Therefore, Eq.~(\ref {Perturbed-Hartry-Fock-eqs2}) has
a group of dynamic symmetry, which algebra is \textsf
{so(3)~х~so(3)} $ \sim $\textsf {so (4)} if to neglect an exchange
interaction. As is known, a nonrelativistic hydrogen-like atom
possesses such symmetry. Hence, we have proved that to calculate
quasiparticle states in the non-relativistic case it is possible
to use a basis set of states of a nonrelativistic hydrogen-like
atom.

However, for a heavy electron $ \Delta M_n (0) \ge 1 $ according
to the formula (\ref {gap}) we always
have
\begin{equation}
a_n=-\Delta {M}_n (0)/2 \label{heavy-electron}
\end{equation}
and, hence, there does not exist    equation such as Schr\"odinger
one for its describing. From here we conclude that the heavy
electron can not be examined in a nonrelativistic limit.

{\it 
Relativistic case}

Let us generalize  the proposed approach to relativistic case. To
do it we substitute Eqs.~(\ref{Particle-energy}) and
(\ref{heavy-electron}) into 
(\ref{Perturbed-Hartry-Fock-eqs1}) and 
let $ n$  tends to 
$ n\to \infty $:
\begin{eqnarray}
 \hat h^{HF}(r_1)\psi_{n}(k_1r_1) =
\left({\Delta {M}_n (0)\over 2}+\tilde {\tilde \epsilon}_n(k_1)
\right)\psi_{n}(k_1r_1),
\quad n\to \infty.
\label{Quasirelativistic-eq}
\end{eqnarray}
Then, one can assume that the operator ${\partial \over \partial
t}- \hat h^{HF}$ in Eq. 
 (\ref{Quasirelativistic-eq})  is a quasirelativistic hamiltonian
 written in the implicit form in the Hartee - Fock approximation.

From  consideration carried out above it  follows that the desired
relativistic equation of motion should describe a charged
composite system from a pair of particles and have the dynamic
symmetry \textsf {SO(4)}. A spin of given quantum system should be
equal 1 as motion of a hole is a motion of   an electron in
many-particle positively charged matrix.
 In \cite{Gurs&Grush-DocBGUIR2003} the equation of motion of a relativistic charged vector boson has been found and shown, that it describes a relativistic hydrogen-like atom.
The relativistic charged vector-boson appears a composite system
with a corresponding spectrum of masses and in quasirelativistic
limit $n\to \infty $ its energy $E_1 $ is determined by the
expression:
\begin{eqnarray}
 E_1\approx {m\over 2} -{m\gamma ^2 \over 2{n }^2} - {m \gamma ^4\over 8 n ^3}
 { \left({4\over   |k|  }-{3\over   n  }\right)}-
 {m \gamma ^6\over 8 n ^4}
 { \left({3\over   n^2  }-{8\over  n  |k|  }+{4\over  k^2  }\right)}
 +O(\gamma^8).\nonumber \\
\ \label{boson-energy}
\end{eqnarray}
Comparison of right-hand sides of formulas (\ref
{Quasirelativistic-eq}) and (\ref {boson-energy}) yields that $
\Delta M _ {\infty} \equiv \lim _ {n\to \infty} \Delta M_n (0) =m
$ is a rest mass  $m $ of an electron. Hence,
Eq.~(\ref{Quasirelativistic-eq}) is an equation of motion for a
relativistic  electron - hole pair with a reduced mass $\Delta M _
{\infty}/2=m/2 $ which, apparently, is the relativistic charged
vector-boson considered in quasirelativistic limit $n\to \infty $.

\section{Conclusion}
So, the variation method to find interacting quasiparticle states in crystals
was developed. The quasiparticle propagator was constructed by summation over
elements of secondary quantized density matrix. This approach allows us to find
  motion equations of one-electron states of excited atom in crystals.

\end{document}